\documentclass[reprint,aps,prb,amsmath,amssymb]{revtex4-1}

\usepackage{graphicx}
\usepackage{dcolumn}
\usepackage{bm}
\usepackage[USenglish]{babel}
\usepackage{xcolor}

\begin{document}

\title{Self-screening corrections beyond the random-phase approximation: Applications to
band gaps of semiconductors}

\author{Viktor Christiansson}
\affiliation{Department of Physics, University of Fribourg, 1700 Fribourg, Switzerland}
\author{Ferdi Aryasetiawan}
\affiliation{Department of Physics, Division of Mathematical Physics, Lund 
University, Professorsgatan 1, 223 63 Lund, Sweden}

\begin{abstract}
The self-screening error in the random-phase approximation (RPA) and the $GW$ approximation (GWA) is a well-known issue and has received attention in recent years with several methods for a correction being proposed.
We here apply two of these, a self-screening and a so-called ``self-polarization" correction scheme, to model calculations to examine their applicability. 
We also apply an explicit self-screening correction to \textit{ab-initio} calculations of real materials. 
We find indications for the self-polarization scheme to be the more appropriate choice of correction for localized states, and additionally we observe that it suffers from causality violations in the strongly correlated regime. The self-screening correction used in this work on the other hand significantly improves the description in more delocalized states. It provides a notable reduction in the remaining GWA error when calculating the band gaps of several semiconductors, indicating a physical explanation for a part of the remaining discrepancy in one-shot $GW$ compared to experiment, while leaving the localized semicore $d$ states mostly unaffected.
\end{abstract}

\maketitle

\section{\label{Sec:Introduction}Introduction}

One of the most important quantities in many-electron theory is the linear density response function, henceforth referred to as response function, which describes how the charge density in a system of electrons is modified upon application of a time-dependent potential. Although the formal expression for the response function is known, in practice one must resort to approximations in order to compute it for real materials, as is the usual case in many-electron theory. Perhaps the most successful and widely used approximation is the random-phase approximation (RPA), \cite{Pines1963} which is also the basic approximation used in the well-known $GW$ approximation (GWA), \cite{Hedin1965} nowadays routinely used to calculate the quasiparticle dispersion in materials. In the GWA the screened interaction $W$ is computed within RPA starting from the Kohn-Sham states and eigenvalues obtained from a DFT \cite{Hohenberg1964,Kohn1965} calculation.

\begin{figure}[ht!] 
\begin{centering}
\includegraphics[width=0.95\columnwidth]{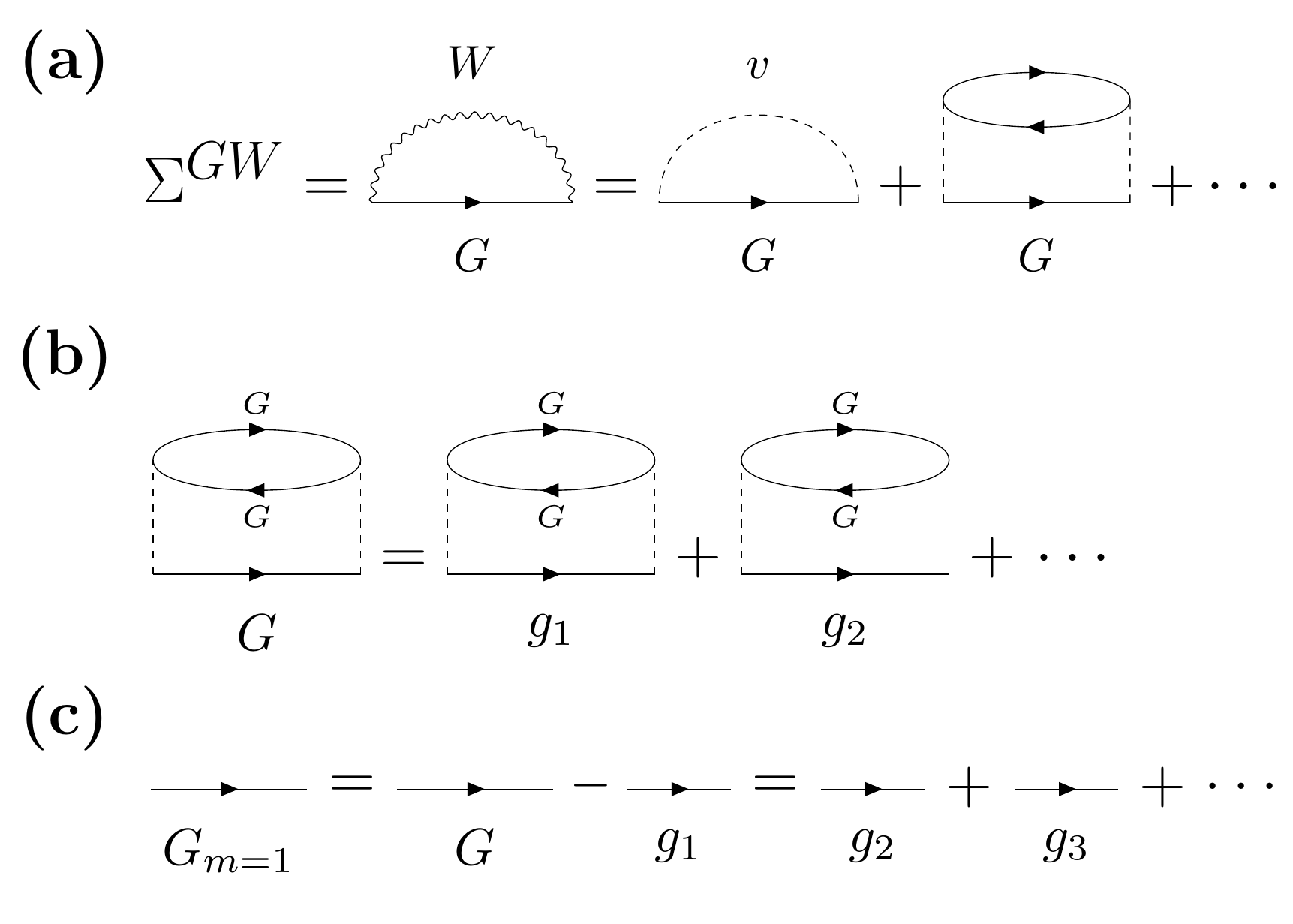}
\par\end{centering}
\caption{(a) Contributions to the self-energy diagram within the GWA. (b) Expansion of the Green's function in orbital Green's functions $g_m$ for one diagram. The presence of the full $G$ in the polarization bubble shows the screening on the propagating electron coming from itself, as discussed in the text.
(c) The Green's function with orbital Green's function $g_{1}$ removed. In the self-screening correction scheme it replaces $G$ in the polarization bubble in the first term of (b).
\label{Fig:FeynmanGW}
}
\end{figure}

A well-known and long-standing problem associated with the RPA is the problem of self-screening. The deficiency can be illustrated most clearly for the case of the hydrogen atom;\cite{Nelson2007} with only one electron in the system, it is not possible to screen the interaction between the electron and an external field since there are no other electrons present, and yet the RPA erroneously yields a non-zero response function. Hence, when the GWA is applied to hydrogen, it yields a correlation self-energy which should be zero. The source of this error arises from a self-screening process inherent in the RPA. As illustrated in Fig.~\ref{Fig:FeynmanGW}(a), the only electron in the system, represented by a Green's function line, should not participate in the polarization bubble; however, in the GWA this self-energy diagram is non-zero.

The self-screening error has received attention in recent years and a few different approaches to treat it have been proposed in the literature.
Romaniello \emph{et al.} \cite{Romaniello2009} used vertex corrections to the self-energy beyond the RPA and applied it to simple model systems. 
Wetherell \emph{et al.} \cite{Wetherell2018} proposed the use of a local potential, based on density functional theory, added to the self-energy, and tested it successfully for simple one-dimensional models. 
In an earlier work by Aryasetiawan \emph{et al.} \cite{Aryasetiawan2012} it was
instead suggested to use a correction based on the introduction of an orbital- and spin-dependent screened interaction, for which it was found that the HOMO-LUMO gap of a hydrogen dimer was correctly reproduced in the weakly to moderately correlated regime.

In the present work, we will employ the schemes proposed in Ref.~\onlinecite{Aryasetiawan2012}
to take the self-screening correction into account.
The idea is that, in a given self-energy diagram, the Green's function line representing the propagation of an electron or a hole of a given orbital is removed from the Green's function appearing in the polarization. In this way, the screening processes associated with the electron or hole of the orbital are eliminated from the total polarization, at the cost of introducing an orbital- and spin-dependent interaction. 
A similar idea can also be applied directly to the response function; instead of removing an orbital line in the Green's function, individual contributions to the polarization associated with an electron-hole excitation are eliminated from the polarization diagram to avoid ``self-polarization".

The aim of this paper is two-fold. First, by using model calculations on a one- and two-orbital Hubbard dimer we gain further insight into the strengths and applicability regimes of the two correction schemes, when applied to response function and self-energy calculations.
Causality violations observed for the self-polarization correction are investigated, and an explanation of the origin is proposed based on the differences and similarities between the two correction schemes. Secondly, we move beyond previous studies of the self-screening error which have only dealt with model calculations, by applying the self-screening correction in fully \textit{ab initio} calculations for a number of semiconductors. The calculated bandgaps and semicore quasiparticle energies are compared to conventional $GW$ calculations and experimental values.

The paper is organized as follows. In Sec.~\ref{Sec:Theory} we introduce the self-screening and self-polarization corrections following the derivation given in Ref.~\onlinecite{Aryasetiawan2012}, and describe a method for making the self-screening correction tractable for the calculation of real materials. In Sec.~\ref{Sec:ResultsModel} and ~\ref{Sec:Causality} we describe our result for the model calculations and discuss the causality considerations, respectively, for the two correction schemes. In Sec.~\ref{Sec:ResultsAbInitio} we present the results from the \textit{ab initio} calculations and compare them to experiments, and finally we discuss and summarize our results in Sec.~\ref{Sec:Conclusions}.

\section{\label{Sec:Theory}Theory and Implementation}

\subsection{Self-screening corrected \emph{GW}}\label{Sec:GW-ss}
In the GWA, \cite{Hedin1965} the self-energy for a spin $\sigma$ is calculated as
\begin{align}
    \Sigma^{GW}_\sigma({\bf r}t, {\bf r'}t') &= i G_\sigma^0({\bf r}t, {\bf r'}t') W({\bf r}'t', {\bf r}t)  \nonumber \\
    &= i \sum_m g_{m\sigma}({\bf r}t, {\bf r'}t')W({\bf r}'t', {\bf r}t),
    \label{Eq:SigmaGW}
\end{align}
where $W$ is the screened interaction, and the non-interacting Green's function $G^0_\sigma$ is decomposed in components of orbital Green's functions $ g_{m\sigma}$. These can, after a transformation to real frequencies, be written in terms of the one-particle orbitals $\varphi_{m\sigma}$ and energies $\varepsilon_{m\sigma}$,
\begin{equation}
    g_{m\sigma}({\bf r}, {\bf r'},\omega) =\frac{\varphi_{m\sigma}({\bf r})\varphi^*_{m\sigma}({\bf r}')}{\omega-\varepsilon_{m\sigma}\pm i\delta},
\end{equation}
with $-$ ($+$) for an occupied (unoccupied) state. 

The self-screening correction scheme for the GWA \cite{Aryasetiawan2012} is based on identifying that an electron in a state $\varphi_{m\sigma}$ takes part in the screening process of itself 
in Eq.~\eqref{Eq:SigmaGW}. 
This occurs when the screened interaction $W$ is calculated within RPA through the inclusion of all propagators $g_{m\sigma}$ in the bubble diagram. 
When calculating the self-energy contribution from an orbital Green's function, the same $g_{m\sigma}$ which is describing the propagating electron is present also in the polarization bubble, as demonstrated in Fig.~\ref{Fig:FeynmanGW}(b). This leads to an unphysical self-screening of the electron.
To remove this self-screening, spin-dependent auxiliary functions $W_{m\sigma}$, calculated within RPA, are defined for each orbital $m$ where the screening from the electron associated with orbital $m$ has been removed. These auxiliary screened interactions are subsequently connected with the corresponding orbital Green's function components in the calculation of the self-energy. 

Schematically, the procedure is formulated as follows. First a Green's function component from orbital $m\sigma$ is removed from the full non-interacting Green's function,
\begin{equation}
    G^0_{m\sigma}=G^0_{\sigma}-g_{m\sigma}.
\end{equation}
A new polarization function for each state $m\sigma$ can then be calculated within RPA with the bubble diagram not including the screening from the $g_{m\sigma}$ line
\begin{equation}\label{Eq:P_ssc}
    P^0_{m\sigma}=-i \left[G^0_{m\sigma}G^0_{m\sigma}+G^0_{-\sigma}G^0_{-\sigma}\right],
\end{equation}
Note that the full contribution to the polarization from the other spin channel is kept, as an electron with opposite spin in orbital $m$ can participate in the screening process. Thereafter, corresponding auxiliary response functions can be obtained as usual through
\begin{equation}
    R_{m\sigma}=P^0_{m\sigma}+P^0_{m\sigma}vR_{m\sigma},
\end{equation}
and finally the orbital- and spin-dependent screened interactions with the self-screening removed are obtained from
\begin{equation}
    W_{m\sigma}=v+vR_{m\sigma}v,
\end{equation}
with $v$ being the bare interaction.
The final step to eliminate the unphysical screening from $g_{m\sigma}$ of itself in the self-energy diagram is by associating in Eq.~\eqref{Eq:SigmaGW} a $g_{m\sigma}$ to its corresponding $W_{m\sigma}$. This reflects the fact that an electron associated with orbital $m\sigma$ experiences a screened interaction $W_{m\sigma}$ in which the contribution to the screening coming from an electron in the same state now has been removed. The corrected self-energy then finally takes the form:
\begin{equation}\label{Eq:Sigma_ssc}
    \Sigma^{GW\textrm{-ss}}_\sigma=i\sum_m g_{m\sigma}W_{m\sigma}.
\end{equation}
It has been shown that this self-screening correction scheme partially corresponds to including exchange diagram vertex corrections to the self-energy.\cite{Aryasetiawan2012}

\subsection{Self-polarization corrected RPA and \emph{GW}}\label{Sec:RPA-sp}

Another view on the self-screening error proposed in Ref.~\onlinecite{Aryasetiawan2012}
is to consider it as arising from an electron-hole excitation, or a dipole.
This so-called self-polarization (to distinguish it from the approach in the previous section), will be outlined next.

To see where the self-polarization error would occur, we decompose the polarization in terms of the individual electron-hole excitations of the system
\begin{equation}
    P^0({\bf r},{\bf r}',\omega)=\sum_\alpha p_\alpha({\bf r},{\bf r}',\omega).
\end{equation}
\begin{align}
    p_\alpha({\bf r},{\bf r}',\omega)=&\frac{\varphi_{m\sigma}({\bf r})\varphi_{n\sigma}^*({\bf r})\varphi_{m\sigma}^*({\bf r}')\varphi_{n\sigma}({\bf r}')}{\omega-\varepsilon_m+\varepsilon_n+i\delta} \nonumber \\
    &-\frac{\varphi_{m\sigma}({\bf r}')\varphi_{n\sigma}^*({\bf r}')\varphi_{m\sigma}^*({\bf r})\varphi_{n\sigma}({\bf r})}{\omega+\varepsilon_m-\varepsilon_n-i\delta},
\end{align}
with the total polarization being the sum over all such excitations.
$\alpha$ is here a combined index of the occupied state $n$ and unoccupied state $m$, as well as the spin $\sigma$. Similar to the discussion in the previous section where an electron screened itself through $R$ (and hence $W$), the given excitation $p_\alpha$ is now involved in the screening of itself.
This can be seen by recognizing that $P^0$ 
appearing in the screening function $[ 1-P^0v ]^{-1}$ includes the excitation $\alpha$ and
hence self-screens the polarization $p_\alpha$:
\begin{equation}\label{Eq:RPAeq}
    R=\left[ 1-P^0 v \right]^{-1}P^0=\sum_\alpha[1-\sum_\beta p_\beta v]^{-1}p_\alpha.
\end{equation}

As before, a new polarization function can be introduced to remedy this,
\begin{equation}
    P_\alpha=P^0-p_\alpha,
\end{equation}
with the excitation $\alpha$ removed. By replacing the full polarization $P^0$ in the screening function $[ 1-P^0v ]^{-1}$ in Eq.~\eqref{Eq:RPAeq} with $P_\alpha$ for each excitation, a new response function corrected for the self-polarization is obtained:
\begin{equation}
    R^{\textrm{RPA\,-\,sp}}=\sum_\alpha \left[ 1- P_\alpha v \right]^{-1}p_\alpha
    \label{Eq:RPAsp}
\end{equation}
With this corrected response function, the calculation of the screened interaction and self-energy then follows as in the usual GWA. Also this form of a self-screening correction can be regarded as an approximate inclusion of higher order exchange diagrams.\cite{Aryasetiawan2012}

\subsection{Self-screening active-space approximation}\label{Sec:Approx method}

In order to apply the self-screening correction in Sec.~\ref{Sec:GW-ss} to \textit{ab initio} calculations for real materials, where in principle the knowledge of an auxiliary screened interaction would be needed for all states $m\sigma$, some level of approximation is necessary to make the method numerically advantageous.
This becomes clear as a large number of unoccupied bands can be required for convergence (as is the case for, e.g., wurtzite ZnO, see Refs.~\onlinecite{Shih2010,Friedrich2011,Friedrich2011Erratum}). Calculating a screened interaction for each band would quickly become highly impractical from a computational perspective. 
We therefore propose to use an approximate form of the method where the self-screening correction is applied only to states in a smaller subspace of the full band structure around the region of interest. We denote this by $\mathcal{S}$.
The self-energy in Eq. \eqref{Eq:Sigma_ssc} is then modified as, separating out the unaffected exchange contribution $\Sigma_\sigma^{x}$ explicitly,
\begin{equation}\label{Eq:Sigma_ssc_approx}
    \Sigma_\sigma=\Sigma_\sigma^{x} + \sum_{m}g_{m\sigma}{\tilde W}^{c}_{m\sigma}
\end{equation}
where the self-screening correction is applied in the calculation of the correlation part of the screened interaction ${\tilde W}^{c}_{m\sigma}$, only if the band $m$ belongs to $\mathcal{S}$. Formally, the screened interaction in the sum in Eq.~\eqref{Eq:Sigma_ssc_approx} then becomes
\begin{equation}
    {\tilde W}^{c}_{m\sigma}=
\begin{cases}
    W^{c}_{m\sigma} \,\,\, \textrm{if } m\in \mathcal{S} \\
    W^{c} \,\,\,\,\,\,\, \textrm{if } m \notin \mathcal{S} 
\end{cases}
.
\end{equation}
($W^{c}$ here indicates that the correlation part of the screened interaction is 
otherwise calculated using RPA within the conventional GWA when $m$ does not belong to the
subspace $\mathcal{S}$.)
This approach effectively turns the size of the chosen subspace into a controllable convergence parameter.

We have implemented the scheme described in this section in the \emph{GW} code \emph{SPEX}. \cite{Friedrich2010}
The preceding DFT calculations in the local density approximation\cite{Hohenberg1964} (LDA) were performed using the full-potential linearized augmented plane-wave (FLAPW) code \emph{FLEUR} \cite{Fleurcode}. The resulting 
Kohn-Sham eigenfunctions and eigenvalues are then used to construct the starting $G^0$ for the one-shot \emph{GW} calculations.
To check for convergence in the studied properties within our scheme, the subspace $\mathcal{S}$ was systematically increased as described in Sec.~\ref{Sec:ResultsAbInitio}.

\section{Results and discussion}

\subsection{Model calculations\label{Sec:ResultsModel}}

\subsubsection{The Hubbard dimer}

In the previous work, \cite{Aryasetiawan2012} the self-screening correction was compared to the GWA and exact results for a simple one-orbital Hubbard dimer model calculation, whereas the effect of the self-polarization correction has not been investigated previously.
There it was found that the self-screening correction correctly describes the HOMO-LUMO gap for the Hubbard dimer with one orbital per site and two electrons (half-filled)
in the limit of small $(U_0-U_1)/2t$, while $GW$ predicts a twice too large contribution from the correlation part of the self-energy. 
This result indicates that the self-screening correction ($GW^{-\textrm{ss}}$) corrects the $GW$ deficiency in the weakly-to-moderately correlated regime for this simple model.
For the self-polarization correction ($GW^{-\textrm{sp}}$) an equivalent derivation gives the gap
\begin{equation}
    \Delta E^{GW\textrm{-sp}}=2t+U_1+\frac{2t}{\Delta_{sp}}\frac{(U_0-U_1)^2}{2t + \Delta_{sp}},
\end{equation}
which, in the limit of small $(U_0-U_1)/2t$, yields the same (incorrect) overestimation as in conventional $GW$. $U_0$ and $U_1$ are the on- and intersite interactions respectively, $t$ the hopping parameter, and the excitation energy is $\Delta_{sp}~=~\sqrt{4t^2+2t(U_0-U_1)}$.
For the Hubbard dimer, the intersite interaction can actually be absorbed into the
onsite one but we have kept it to see its effects more explicitly.

Taking now instead the other limiting case of small hopping $t$, which physically can be interpreted as increasing the separation between the two sites moving towards an isolated atomic picture, the exact gap reduces to $\lim_{t\to0}\Delta E^{\textrm{exact}}=U_0$. We note that this becomes independent of the intersite Coulomb interaction $U_1$. Both the $GW$ and $GW^\textrm{-ss}$ results, however, reduce to $(U_0+U_1)/2$, whereas intriguingly the $GW^\textrm{-sp}$ correction correctly follows the exact result as $\lim_{t\to0}\Delta E^{GW\textrm{-sp}}~=~U_0$. 
Even when the intersite interaction is negligible, the gap predicted by $GW$ and $GW^\textrm{-ss}$ is still found to be half that of the exact and $GW^\textrm{-sp}$ results. Similarly, in the regime of large $U_0$ the self-polarization correction predicts the gap closest to the exact result. 
Taken together, it is indicative for that the two correction schemes are complementary and valid in different regimes.

\begin{figure}[ht!] 
\begin{centering}
\includegraphics[width=0.45\textwidth]{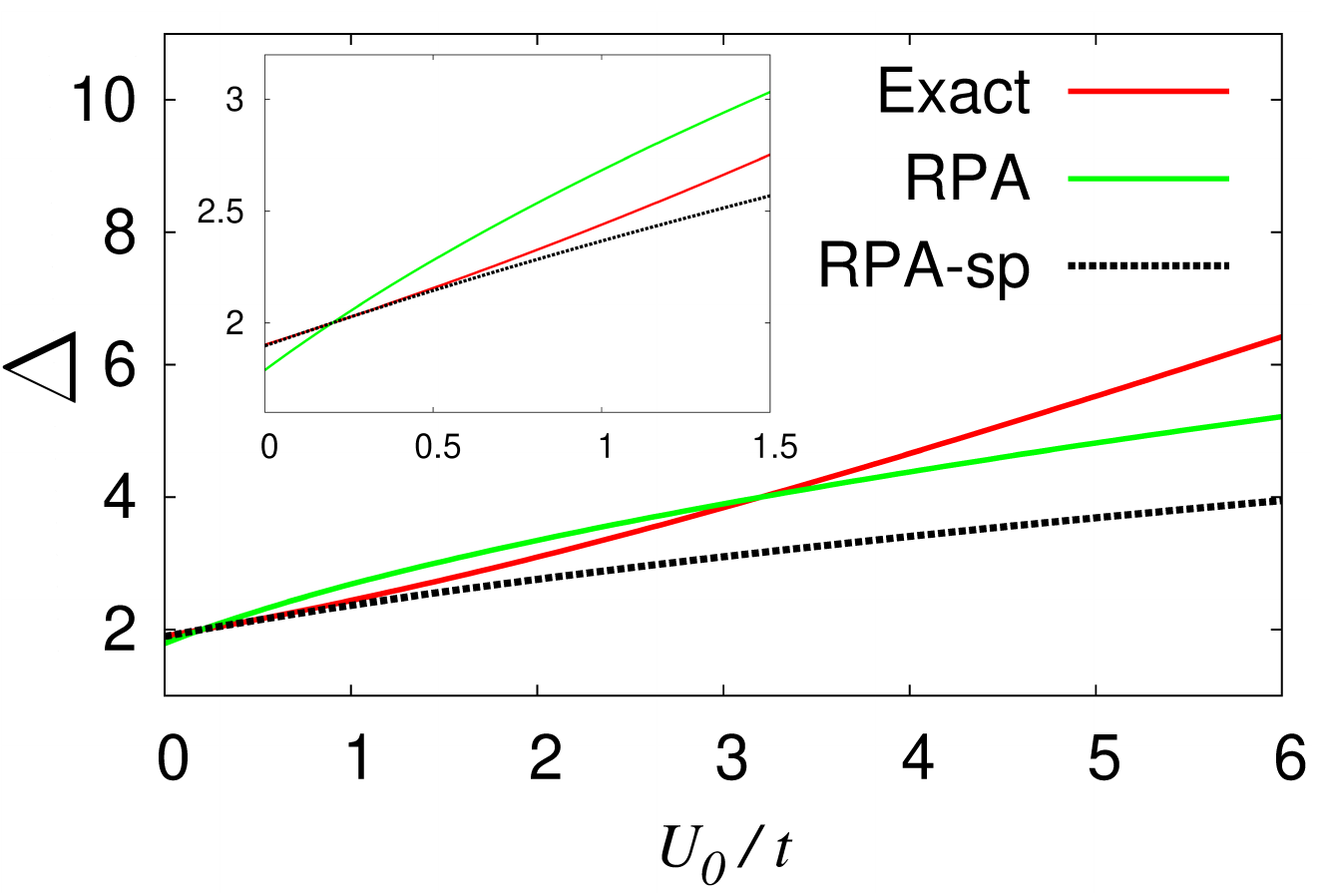}
\par\end{centering}
\caption{Exact, RPA, and self-polarization corrected RPA excitation energies for the one-orbital Hubbard dimer in units of $t=1$ for a fixed $U_1=0.2$. The inset shows the small interaction limit.
\label{Fig:NparticleExc}
}
\end{figure}

Two further effects coming from the correction schemes not previously discussed are how they change the excitation energies of the $N$-electron system, obtained from the response function $R$, and the $(N\pm 1)$-electron excitation energies obtained from the renormalized Green's function.
The two-electron excitation energy is found to be in almost perfect agreement with the exact result for both of the correction schemes
\footnote{The response functions in the self-screening correction are merely auxiliary quantities, and for the one-orbital dimer they are simply related to the self-polarization response as $R^\textrm{ss}_{m\sigma}=R^\textrm{sp}/2$. The predicted $N$-particle excitation energy is therefore the same.}
in the regime $(U_0-U_1)/t<1$, while the excitation energy found from the conventional RPA is both qualitatively and quantitatively incorrect. This is indicated in the inset of Fig.~\ref{Fig:NparticleExc}.

With increasing interaction strength all methods deviate from the exact excitation energies. At $U_0/t\gtrsim 1.5$ the energies within RPA is in a better agreement compared to the correction schemes, although with the same qualitatively wrong large interaction behaviour. This more ``accidental" improvement in RPA at large $U_0/t$ is to be compared with the good agreement in the low-to-moderate interaction strength regime for the two correction schemes.

The positions of the quasiparticle peaks in the renormalized Green's functions (not shown) are close to the exact result for all methods at low-to-moderate interaction strengths, however, both correction schemes are found to approximately halve the incorrect overestimation of the satellite positions in $GW$. In the large interaction regime the self-polarization correction predicts the peak closest to the exact result, while simultaneously both $GW^\textrm{-ss}$ and $GW^\textrm{-sp}$ maintain their improved description of the satellites.

\subsubsection{Two-orbital Hubbard dimer}

We next extend the model calculations to a dimer model with two orbitals per site, 
with a Hamiltonian of the form
\begin{equation}
    H=-\sum_{ij\alpha\beta} t_{i\alpha,j\beta}\hat{c}^\dagger_{i\alpha\sigma}\hat{c}_{j\beta\sigma} + \sum_{i\alpha} U_\alpha \hat{c}^\dagger_{i\alpha\uparrow}\hat{c}_{i\alpha\uparrow}\hat{c}^\dagger_{i\alpha\downarrow}\hat{c}_{\alpha\downarrow}
\end{equation}
where $\hat{c}^\dagger_{i\alpha\sigma}$ ($\hat{c}_{i\alpha\sigma}$) is the creation (annihilation) operator of an electron with spin $\sigma$ at site $i$ in orbital $\alpha$, and $t_{i\alpha,j\beta}$ the hopping parameters. We have restricted the interaction $U_\alpha=U_0$ to be non-zero only for the lower orbital $\alpha=1$. For simplicity, we here only vary the interaction strength while fixing the hopping parameters in units of the inter-site hopping $t_{11,21}=t_{21,11}=t=1$, with the specific choice of $t_{12,22}=t_{22,12}=0.5t$, $t_{11,22}=t_{22,11}=0.2t$, and an orbital energy separation $t_{i1,i1}~-~t_{i2,i2}~=~2t$ used for the figures in this section (the intra-site hopping $t_{i\alpha,i\beta}$, $\alpha\neq\beta$, is absorbed as a shift in the orbital energy separation).
We have checked that other choices of parameters do not affect the discussions.

\begin{figure}[t] 
\begin{centering}
\includegraphics[width=0.48\textwidth]{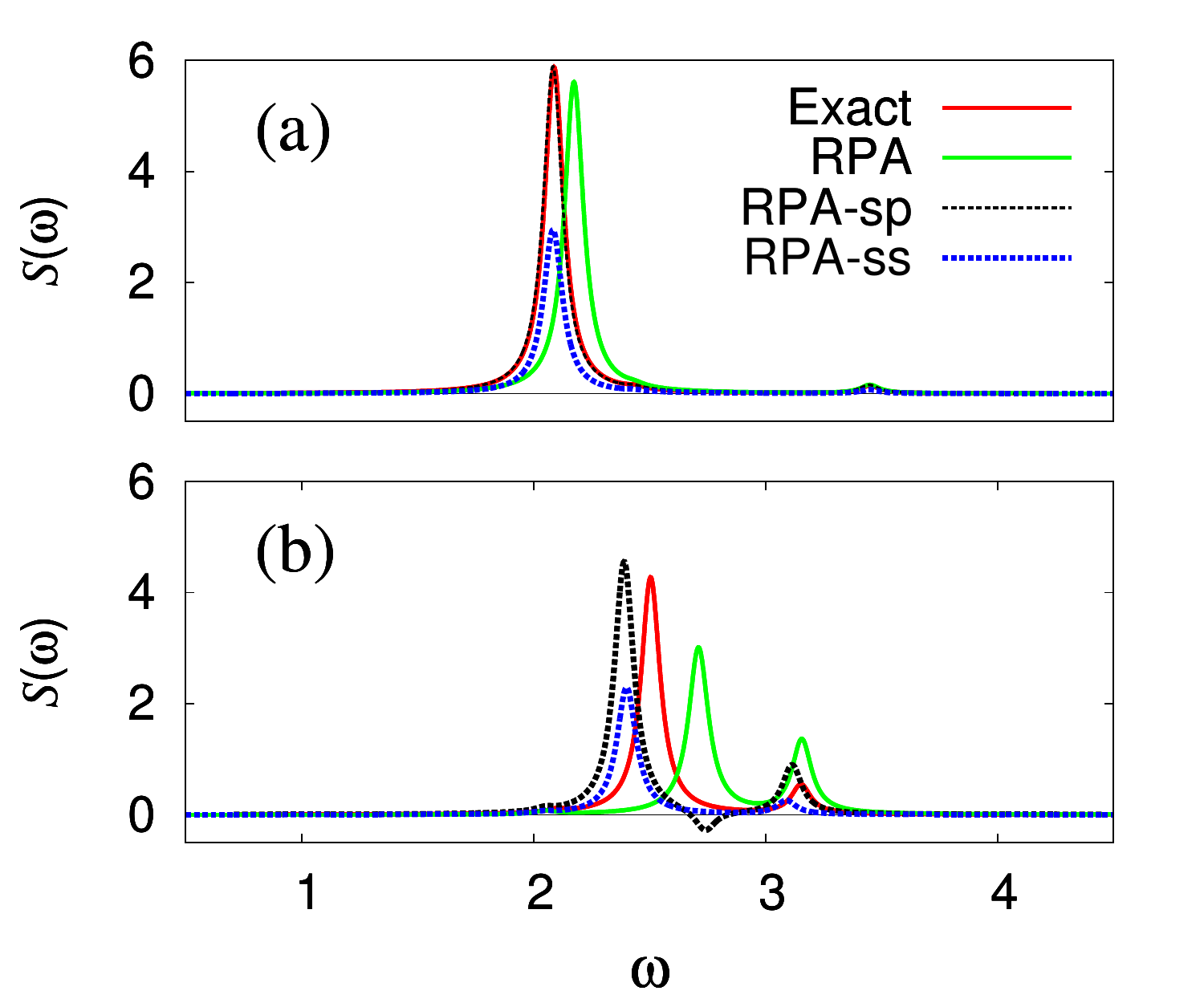}
\par\end{centering}
\caption{Diagonal components of the exact, RPA, and self-polarization corrected RPA spectral response functions for the two-orbital Hubbard dimer. Also shown is one auxiliary response function calculated using the self-screening correction. a) is calculated for an interaction $U_0/t=0.2$, and b) for $U_0/t=1.0$.
\label{Fig:Response}
}
\end{figure}

In the regime of lower interaction strengths ($U_0/t < 1$) the correction schemes provide an excellent agreement with the exact result for the excitation energies obtained from the response function as indicated in Fig.~\ref{Fig:Response}. At larger interaction strengths, as would be expected, all the RPA based schemes fail and deviate significantly from the exact results. 
We can at this point also note an important difference between the two correction schemes: for moderate interaction strengths ($U_0/t \gtrsim 1$), causality violations are seen to emerge from the self-polarization correction as indicated in Fig.~\ref{Fig:Response} (b), where a partially negative spectral response function is found. These were not present in the single-orbital model and the origin will be further discussed in Sec.~\ref{Sec:Causality}.

To conclude the model calculations we will briefly discuss the renormalized Green's function. The quasiparticle peaks for smaller interaction strengths are only marginally renormalized, while the satellite features in the spectral functions also for the two-orbital model are found to be significantly improved in both correction schemes over those calculated within $GW$. The errors in the satellite positions are again approximately halved for small to intermediate $U_0/t$, whereas all $GW$ based methods are worsened significantly for larger interactions. The non-causality occurring in the self-polarization correction scheme is also observed for the spectral function.

\subsection{Causality considerations} \label{Sec:Causality}

As we noted in the previous section, the self-polarization correction can produce non-causal features in the response function, specifically in parameter regimes where the electrons are more localized, leading to similar issues also in the renormalized Green's function. In contrast, we have not observed any non-causality in any of the calculations in this work based on the self-screening correction. As the corrections in spirit are very similar, we propose a simple picture for
the origin by focusing on the effect of the corrections from the perspective of a single occupied state $\varphi_{m\sigma}$.

The correction in the self-screening scheme amounts to the state $\varphi_{m\sigma}$ being completely removed in the screening of itself, as discussed previously, by virtue of it being completely removed from the polarization propagator $P_{m\sigma}^0$ in Eq.~\eqref{Eq:P_ssc}. On the other hand, in the self-polarization scheme what is removed from the screening through the polarization $P_\alpha$ is an electron-hole excitation with the occupied state $\varphi_{m\sigma}$ and an unoccupied state $\varphi_{n\sigma}$. Through the calculation of $R$ in Eq.~\eqref{Eq:RPAsp} it is then clear that it is this excitation which is effectively prohibited from screening itself, not the physical states involved. This in turn leads to only a partial removal of the physical self-screening coming from the electron in state $\varphi_{m\sigma}$, as it is still involved in the screening of itself through another excitation involving the unoccupied state $\varphi_{n'\sigma}$.

To make the physical picture clearer, we can draw a comparison to the model dimer calculations. In the case of the single-orbital dimer all results remain causal, since there exists only one unoccupied (antibonding) state that the bonding state, occupied by two electrons of opposite spin, can be excited to. When this excitation is removed, the only state involved in the screening process is for the electron with opposite spin. 
If the system instead has three unoccupied orbitals into which the electrons can be excited, as is the case for the two-orbital dimer, the issue becomes clear;
whereas in the self-screening corrected (auxiliary) polarization $P_{m\sigma}^{-\textrm{ss}}$ all instances of the occupied state $\varphi_{m\sigma}$ have been removed, 
the self-polarization scheme provides three different $P_\alpha$, all being excitations from the occupied $\varphi_{m\sigma}$ to the three unoccupied states. In the summation over excitations in Eq.~\eqref{Eq:RPAsp},
$$
R^{\textrm{RPA\,-\,sp}}=\sum_\alpha \left[ 1- P_\alpha v \right]^{-1}p_\alpha
$$
the occupied $\varphi_{m\sigma}$ is then only partially removed for each term, as the other excitations retain its presence in $P_\alpha$.
We therefore believe we can attribute the non-causality observed for the self-polarization correction to be due to a partial removal of the physical state from screening itself. 
This observation could be used as a starting point to develop a causal self-polarization scheme, and we plan to investigate this in a future work.

\subsection{\emph{Ab initio} calculations \label{Sec:ResultsAbInitio}}

\begin{table*}[t]
\setlength{\tabcolsep}{15pt}
\renewcommand{\arraystretch}{1.5}
    \centering
        \caption{Band gaps in eV calculated within LDA, $GW$, and self-screening corrected $GW$. The self-screening correction was applied with 20 and 40 bands respectively included in the subspace $\mathcal{S}$. The experimental values are taken from Refs.~\onlinecite{Madelung2002,Chiang1989,Madelung1999} as indicated. The values in parenthesis for Ge and Si give the direct gap at the $\Gamma$-point, and the values in square brackets for ZnO show the extrapolated band gaps discussed in the text.}\label{Table:bandgap}
    \begin{tabular}{|c||c|c|c|c|c|}
    \hline
           & LDA & $GW$ & $GW^{\textrm{-ss}}$ (20) & $GW^{\textrm{-ss}}$ (40) & Exp. \\
         \hline
         \hline
        Ge & - & 0.70 (0.80) & 0.79 (0.89) & 0.79 (0.89) & 0.74 (0.90) \cite{Madelung2002} \\
        \hline
        Si & 0.47 (2.52) & 1.00 (3.16) & 1.12 (3.25) & 1.12 (3.25) & 1.17 (3.37) \cite{Chiang1989} \\
        \hline
        InP & 0.47 & 1.24 & 1.36 & 1.36 & 1.42 \cite{Madelung2002} \\
        \hline
        GaAs & 0.30  & 1.38 & 1.50 & 1.50 & 1.52 \cite{Madelung2002} \\
         \hline
        CdTe & 0.51 & 1.43 & 1.55 & 1.55 & 1.61 \cite{Madelung1999} \\
        \hline
        ZnSe & 1.04 & 2.42 & 2.60 & 2.60 & 2.82 \cite{Madelung1999} \\
        \hline
         ZnO & 0.76 &  2.41 [2.79] & 2.56 [2.94] & 2.57 [2.95] & 3.44 \cite{Madelung1999} \\
        \hline
    \end{tabular}
\end{table*}

Next we move beyond the model calculations of the previous sections to test the reliability of the self-screening correction also for predicting properties of real materials. We have carried out \emph{ab initio} calculations for several semiconductors and focused on two well-characterized problems for a ``one-shot" $GW$ ($G^0W^0$) calculation: the band gap and the 
position of the more localized semicore states. For these calculations we used the self-screening correction with the implementation as described in Sec.~\ref{Sec:Approx method}.
The band gap is known to be significantly improved within the GWA compared to DFT calculations, although there is still a slight underestimation for many materials.\cite{Hybertsen1986,Hott1991,Schilfgaarde2006}
For several semiconductors and insulators this can be improved by introducing a limited form of self-consistency in the so-called quasiparticle self-consistent $GW$.\cite{Schilfgaarde2006}
Semicore states calculated within the GWA, albeit improved over those obtained from LDA calculations, are found to still be located too high in energy. \cite{Aryasetiawan1996}

The quasiparticle energies were obtained by finding the solution to the quasiparticle equation \cite{Hybertsen1986}
\begin{equation}
    E^{QP}_{n{\bf k}}=\varepsilon^{DFT}_{n{\bf k}} + \Sigma_{n{\bf k}}(E^{QP}_{n{\bf k}}) - V^{XC}_{n{\bf k}}
\end{equation}
for a state $n{\bf k}$ where the self-energy corrects the exchange-correlation potential $V^{XC}$ contained in the one-particle energies $\varepsilon^{DFT}_{n{\bf k}}$. 
To benchmark the correction we have applied it to several semiconductors with varying band gaps: Si, GaAs, Ge, CdTe, InP, ZnSe, and wurtzite ZnO, where the six latter ones are known to also have high-lying $d$ semicore states. A $8 \times 8 \times 8$ ${\bf k}$-grid was used for the calculations ($12 \times 12 \times 12$ in the case of Si to more accurately estimate the location of the indirect bandgap).
Both the $GW$ and $GW^\textrm{-ss}$ calculations are performed within $G^0W^0$, where the non-interacting Green's function is obtained directly from a preceding DFT calculation, and the Green's function is not subsequently updated in further iterations.

The band gaps calculated for the semiconductors within LDA, $GW$, and $GW^\textrm{-ss}$ are listed in Table~\ref{Table:bandgap} and compared with the experimental values. 
To test the convergence of the approximate treatment of the self-screening correction, we carried out convergence tests by systematically increasing the active region $\mathcal{S}$.
We find rapidly converging results already with the 20 lowest bands included in the subset treated with the correction, counting from the high-lying semicore $d$ states. No change is observed when $\mathcal{S}$ is further doubled in size to the 40 lowest bands. This important result we believe justifies the approximate treatment, and shows the method to be numerically tractable for realistic material calculations.

For all semiconductors, we find the band gaps which are underestimated within $GW$ to be increased when the correction scheme is applied, bringing them to a closer agreement with the experimental values. The only exception is Ge which is predicted to be a metal in the LDA calculation, where the band gap is already relatively well described by $GW$. The use of the correction instead leads to a slight overestimation of the gap, with the error being comparable to the one in conventional $GW$. The observed trend of increasing gaps is in agreement with the expectation that a reduction in screening would bring the situation more towards a Hartree-Fock picture (here the self-energy contains only the exchange part, $\Sigma^x$, neglecting screening effects), where it is well-known that too large band gaps are found. \cite{Svane1987,Onida2002,Hott1991}

It has previously been shown \cite{Shih2010,Friedrich2011,Friedrich2011Erratum} that a very large number of unoccupied bands is required to converge the band gap for wurtzite ZnO.
Assuming our approximate treatment of the self-screening correction is fully converged within the chosen $\mathcal{S}$ subset of bands, by increasing the number of untreated unoccupied bands a similar situation as for a usual $GW$ calculation should occur. 
We therefore follow Friedrich \emph{et. al.} in Ref.~\onlinecite{Friedrich2011} and use a hyperbolical fit for $N$ bands of the form 
\begin{equation}
    E_\textrm{gap}(N)=\frac{a}{N-N_0}+b,
\end{equation}
with parameters $a$, $b$, and $N_0$ extracted from the corrected fit in Ref.~\onlinecite{Friedrich2011Erratum}. As they similarly used the FLAPW method in their work, we believe our results should be comparable. Indeed, by extrapolating the hyperbolical fit shifted to match our values calculated at $N=500$ bands, we find a good agreement between our $GW$ asymptote at 2.79 eV and the previously reported value of the band gap at 2.83 eV.\cite{Friedrich2011Erratum}
\begin{table*}[t]
\setlength{\tabcolsep}{15pt} 
\renewcommand{\arraystretch}{1.5} 
    \centering
        \caption{Position of the center-of-weight of the $d$ semicore states in eV, 
        relative to the valence band maximum calculated within LDA, $GW$, and self-screening corrected $GW$. The self-screening correction was applied with 20 and 40 bands respectively, and additionally with only the semicore states included in the subspace $\mathcal{S}$. The experimental positions of the semicore states are taken from Ref.~\onlinecite{Chiang1989}, averaged over the reported $d_{3/2}$ and $d_{5/2}$ energies.}    \label{Table:semicoreVBM}
    \begin{tabular}{|c||c|c|c|c|c|c|}
    \hline
         & LDA & $GW$ & $GW^{\textrm{-ss}}$ ($ d$) & $GW^{\textrm{-ss}}$ (20) & $GW^{\textrm{-ss}}$ (40) & Exp. \\
        \hline
        \hline
        Ge & 24.7 & 27.0 & 27.0 & 27.0 & 27.0 & 29.6 \\
        \hline
        InP & 14.2 & 15.5 & 15.5 & 15.5 & 15.5 & 17.3 \\
        \hline
        GaAs & 14.8 & 16.5 & 16.7 & 16.6 & 16.6 & 18.8 \\
        \hline
        CdTe & 8.0 & 8.9 & 9.00 & 8.9 & 8.9 & 10.5 \\
        \hline
        ZnSe & 6.4 & 7.4 & 7.5 & 7.4 & 7.4 & 9.3 \\
        \hline
        ZnO & 5.1 & 5.8 & 5.8 & 5.7  & 5.7 & 8.9 \\
        \hline
    \end{tabular}
\end{table*}
The similarly extrapolated gap obtained from the self-screening corrected results brings our final gap to a better agreement at 2.95 eV. The bandgaps resulting from this fit are indicated in Table~\ref{Table:bandgap}.

In Table~\ref{Table:semicoreVBM} we finally present our calculated center-of-weight values of the $d$ semicore states, averaged over bands and the Brillouin zone, in GaAs, Ge, InP, CdTe, ZnSe, and ZnO relative to the valence band maximum. We observe a very minor improvement compared to the $GW$ values for some materials when only treating the semicore states with the self-screening correction. 
This small change is cancelled and even reversed when increasing the number of bands treated. On closer inspection, however, this can be explained by the reported widening of the band gap in Table~\ref{Table:bandgap}.
Instead, comparing the position with the conduction band minimum, or the center of the gap, conversely shows a slight further improvement in the energy position. 
This change is, however, negligible, and shows that the localized semicore $d$ states remain mostly unaffected by the self-screening correction.

\section{Conclusions \label{Sec:Conclusions}}

We have applied two recently proposed self-screening corrections to RPA (self-polarization) and the GWA (self-screening) in model systems to further investigate the effects and regions of validity of the schemes compared with the conventional methods. We have furthermore employed one of the schemes, the self-screening correction, in \emph{ab-initio} calculations for quantitative predictions of a number of semiconductors, specifically studying its effect on the band gaps and the position of the semicore $d$ states. In order to apply the correction to realistic materials calculations we proposed an approximate scheme, where we introduce an ``active-space" with only a subset of the full underlying DFT band structure used in the construction of $G^0$ being treated with the correction. We have shown that this approximate treatment quickly converges in the number of required bands for the properties studied, justifying it as a numerically advantageous method.

From our model calculations we can draw two clear conclusions regarding the differences between the correction schemes:
\begin{itemize}
    \item[1.] The self-polarization scheme corrects the HOMO-LUMO gap in the localized (strongly correlated) regime, whereas for low interaction strengths it reduces to the $GW$ result. This can be compared with the self-screening correction, which instead reproduces the exact result in the low-to-moderate interaction (delocalized) regime as shown earlier in Ref.~\onlinecite{Aryasetiawan2012}.
    
    \item[2.] The self-polarization scheme suffers from causality violations, not observed for the self-screening correction. We have proposed an origin of this issue from a picture based on an occupied physical state, where the self-screening is only partly removed from the different excitations in the system.
\end{itemize}
The two schemes also display some similarities. Specifically, we find that the excitation energies are well described for both the self-screening and self-polarization schemes in the low-to-moderate interaction strength regime. Further, we observe that both correction schemes predict an improved position of the satellite features in the spectral function, as compared with the conventional GWA. Taken together, this indicates that the two formulations could be considered complementary in more delocalized (self-screening) and localized (self-polarization) cases.

The \emph{ab-initio} results demonstrate that the error in the band gap is decreased notably when the self-screening correction is applied, predicting values within only a few percent from experiments in most cases. That the more delocalized $sp$ states involved around the band gap have a large part of the remaining error in the GWA compared to experiment removed agrees with what we would expect based on the model calculations. Similarly, the more localized semicore states show no noticeable changes from the correction, strengthening our supposition that the self-screening correction is more suitable for improving the description of delocalized states. Based on these findings, we believe that it would be of interest to apply also the self-polarization scheme in realistic calculations of materials; specifically, the more localized $d$ semicore states could be expected to show an improvement. However, the issue of non-causality would first have to be addressed. The proposed origin of this issue, an incomplete removal of the self-screening from a given physical state, could provide a starting point for an improved scheme.

To summarize, it is well-known that RPA and $GW$ based methods overall work well for more delocalized systems. 
Our calculations for both models and real materials indicate that a significant part of the remaining error in a conventional GWA calculation for such delocalized states can be reduced by considering a correction of the self-screening error coming from the RPA.
While self-screening can be expected to also be significant for a localized state, our model calculations and \emph{ab-initio} predictions of $d$ semicore state positions indicate that some form of the so-called self-polarization correction could prove more suitable in this regime.
To further verify the usefulness of the self-screening correction, investigations of additional classes of materials would be of interest.

\begin{acknowledgments}

V.C. acknowledges support from the European Research Council through ERC Consolidator Grant No. 724103. F. A. gratefully acknowledges financial support from the Knut and Alice Wallenberg
(KAW) Foundation Grant No. 2017.0061 and the Swedish Research Council, Vetenskapsrådet (VR)
Grant. No. 2021-04498\_3.
The calculations were performed on the Beo05 cluster at the University of Fribourg.
\end{acknowledgments}

\appendix

\bibliography{main}%

\end{document}